\documentclass[9pt]{article}
\usepackage{spconf}
\usepackage{flushend,cite}
\usepackage{amsmath,amssymb,amsfonts}
\usepackage{algorithmic}
\usepackage{amsmath}
\usepackage[export]{adjustbox}
\usepackage{graphicx}
\usepackage{subcaption}
\usepackage{textcomp}
\usepackage{xcolor}

\def\BibTeX{{\rm B\kern-.05em{\sc i\kern-.025em b}\kern-.08em
    T\kern-.1667em\lower.7ex\hbox{E}\kern-.125emX}}
\usepackage[hyphens]{url}
\usepackage{hyperref}
\usepackage{booktabs}
\usepackage{tikz}
\usetikzlibrary{positioning}
\usetikzlibrary{shapes.geometric}

\name{\begin{tabular}{c}Veronica Wairimu Muriga$^{a,1}$,  Benjamin Rich$^{a}$, Francesco Mauro $^{b}$,\\
\textit{Alessandro Sebastianelli}$^{c}$ \textit{and Silvia Liberata Ullo}$^{b}$\end{tabular}\thanks{
$^{1}$Corresponding author. 
\textit{Email addresses}: wmuriga$@$mit.edu (VWM), brrich$@$mit.edu (BR), f.mauro$@$studenti.unisannio.it (FM), Alessandro.Sebastianelli$@$esa.int (AS), ullo$@$unisannio.it (SLU)}
}

\address{
$^{a}$ Massachusetts Institute of Technology, Boston, USA \\
$^{b}$ Engineering Department, University of Sannio, Benevento, Italy \\
$^{c}$ $\phi$-lab, European Space Agency, Frascati, Italy\\
}

\title{A machine learning approach to long-term drought prediction using Normalized Difference Indices computed on a spatiotemporal dataset }

\begin{document}
\maketitle

\begin{abstract}
\vspace{-0.2cm}

Climate change and increases in drought conditions affect the lives of many and are closely tied to global agricultural output and livestock production. 
This research presents a novel approach utilizing machine learning frameworks for drought prediction around water basins. 
Our method focuses on the next-frame prediction of the Normalized Difference Drought Index (NDDI) by leveraging the recently developed SEN2DWATER database. 
We propose and compare two prediction methods for estimating NDDI values over a specific land area. Our work makes possible proactive measures that can ensure adequate water access for drought-affected communities and sustainable agriculture practices by implementing a proof-of-concept of short and long-term drought prediction of changes in water resources.

\end{abstract}

\begin{keywords}
Climate change, Drought, Sentinel-2, Water Indices, Deep Learning, Late/Early Stage Computation
\end{keywords}
\vspace{-0.35cm}
\section{Introduction}
\vspace{-0.25cm}

Previous studies in remote sensing and drought assessment have led to the development of several indices computed from multispectral and hyperspectral data \cite{koppe2010, Moreira2015}. 
Among these indices, the Normalized Difference Moisture Index (NDMI) \cite{8899794} is valuable for monitoring vegetation moisture levels, while the Normalized Difference Vegetation Index (NDVI) \cite{AFAQ2021101310, rahman2019change} helps assess the presence of live green vegetation in a landscape. 
The Normalized Difference Drought Index (NDDI), which combines information from NDVI and NDMI, has been shown to be particularly useful for mapping drought severity and identifying drought-prone areas at an early stage \cite{Artikanur2022-ns, https://doi.org/10.1029/2006GL029127}. 

Machine learning (ML) has opened new avenues for climate and weather forecasting. Recent advancements have demonstrated the successful application of various neural network architectures to remote sensing data \cite{9769919, 8861302, hunt2022using, lee2020future, shi2015convolutional}. Machine learning techniques have additionally been employed to predict changes in vegetation health, water content, quality, and moisture levels using normalized indices \cite{AFAQ2021101310, water2022, ferchichi2022forecasting}. 
Among various frameworks for predicting such indices, Time Distributed Convolutional Neural Networks (TD-CNNs) show promise in time series forecasting as they combine CNN layers for pattern recognition in images with a Long-Short Term Memory (LSTM) layer capable of identifying changes across time series data (\href{https://machinelearningmastery.com/timedistributed-layer-for-long-short-term-memory-networks-in-python/}{Using TD-CNNs with Keras}).

In this study, we expand on the application of ML methods to normalized difference indices, focusing on NDDI. Previous research has shown that the NDDI, computed using visible, near-infrared, and short-wave infrared channels, exhibits enhanced sensitivity as a drought indicator compared to other indices \cite{https://doi.org/10.1029/2006GL029127}. Therefore, we investigate the effectiveness of NDDI as an early indicator of drought or significant environmental changes using data from European landscapes near water bodies.

We train TD-CNNs on NDVI, NDMI, and NDDI data obtained from SEN2DWATER, a 2-dimensional dataset derived from multispectral Sentinel-2 data spanning six years. 
We explore two approaches for learning and predicting NDDI values, by changing the point at which NDDI computation is performed. 
We present our findings and provide insights into the performance differences between these two methods. 
Furthermore, we discuss potential applications and use cases for NDDI prediction. The code for this work can be found on our (\href{https://github.com/veronicamuriga/USANNIOMIT}{GitHub page}).

\vspace{-0.4cm}
\section{Dataset}
\vspace{-0.25cm}

We utilize the SEN2DWATER dataset (\href{https://github.com/francescomauro1998/SEN2DWATER}{SEN2DWATER: A Novel Multitemporal Dataset and Deep Learning Benchmark For Water Resources Analysis}), a spatiotemporal dataset created from multispectral Sentinel-2 data collected over water bodies from July 2016 to December 2022. SEN2DWATER contains data from all 13 bands of Sentinel-2, making it suitable for our research \cite{Mauro2023}. The dataset was curated by selecting the least cloudy samples from Sentinel-2 within a two-month period, resulting in 39 satellite images per location over a span of six and a half years. Notably, SEN2DWATER primarily comprises spectral data from lakes and rivers in Spain. Therefore, the areas chosen for this study are not highly prone to drought, which could potentially affect the effectiveness and applicability of the NDDI index.   
 
From SEN2DWATER, we acquire landscape images spanning multiple years that capture the spatial and spectral information required to compute the indices employed in our study. The formulas for calculating the NDMI, NDVI, and NDDI indices are presented in Equations 1-3, as described in \cite{https://doi.org/10.1029/2006GL029127}\footnote[2]{Note that the authors of reference \cite{https://doi.org/10.1029/2006GL029127} refer to NDMI as NDWI. To ensure clarity, we have renamed the NDWI equation used in \cite{https://doi.org/10.1029/2006GL029127} to NDMI, as there are two distinct and different definitions for the NDWI index.}.

\vspace{-5mm}
{\small
\begin{align}\label{eq:indices}
\vspace{-1cm}
NDVI &= \frac{NIR (Band 8)-Red (Band 4)}{NIR (Band 8) + Red (Band 4)}\\[2mm]
NDMI &= \frac{NIR (Band 8) - SWIR(Band 11) }{NIR (Band 8) + SWIR(Band 11)}\\[2mm]
NDDI &= \frac{NDVI-NDMI}{NDVI+NDMI}
\end{align}
}

We extract spectral data from SEN2DWATER for several locations in Italy and Spain. This data is used to generate $300px\times 300px$ images for each index, namely NDMI, NDVI, and NDDI, across the entire time series for each location in the dataset. Since all Sentinel-2 bands are captured almost simultaneously, a single image in the time series represents the simultaneous capture of NDVI and NDMI for the same location. Subsequently, these images are divided into 16 separate $64px\times 64px$ images, excluding the remaining data. Each of these three sets of data (NDVI, NDMI, and NDDI) is further divided into training and validation sets. 

\vspace{-0.35cm}
\section{METHODS}
\vspace{-0.25cm}

\subsection{Model}
\vspace{-1mm}
We use a TD-CNN to perform a prediction of the next frame from a time-series sequence of images from various landscapes. A diagram of the TD-CNN architecture is illustrated in Figure \ref{TD-CNN}, along with a simplified illustration of the general workflow used to compute predicted NDDI values.

A TimeDistributed class which applies a layer to every slice of a temporal input is offered by Tensorflow (\href{https://keras.io/api/layers/recurrent_layers/time_distributed/}{Keras: TimeDistributed Layer}). The TimeDistributed layer trains a set of 2D CNN blocks to detect features in a series of images, such that the layer contains information on image characteristics that change from frame to frame. We apply the same 2D convolutional layer to each timestamp independently, such that the same set of weights is applied to all images. TD-CNNs, therefore, give the advantage of saving on space, since the number of parameters does not increase with an increase in the number of images being learned on, while still allowing the model to learn from many images separately with just one layer. 
We apply a Convolutional LSTM layer after the Time-Distributed layers to capture chronological information obtained from the Time Distributed layers (\href{https://levelup.gitconnected.com/hands-on-practice-with-time-distributed-layers-using-tensorflow-c776a5d78e7e}{Hands-On Practice with Time Distributed Layers using Tensorflow}).
CNNs are useful in extracting features from individual frames, while LSTMs are useful for interpreting the extracted features across time steps (\href{https://machinelearningmastery.com/cnn-long-short-term-memory-networks/#:~:text=A%20CNN%20LSTM%20can%20be,the%20features%20across%20time%20steps}{CNN LSTM Networks}). Therefore, TD-CNNs combine the advantages of CNNs with those of LSTMs, as highlighted before.

\begin{center}
\vspace{-.2cm}
\begin{figure}[!ht]
    \includegraphics[width = 8.5cm]{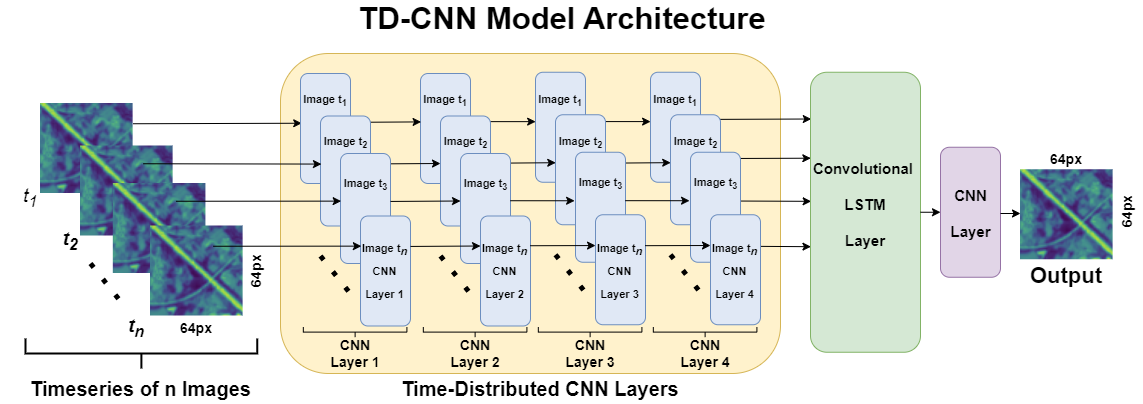}
    \caption{TD-CNN Model Architecture. }
    \label{TD-CNN}
\end{figure}
\vspace{-1cm}
\end{center}

\subsection{NDDI Computation and Prediction}
\begin{figure}[!ht]
    \centering
    \begin{subfigure}[b]{.5\textwidth}
        \caption{Early NDDI Computation}
        \includegraphics[width = 
        \textwidth]{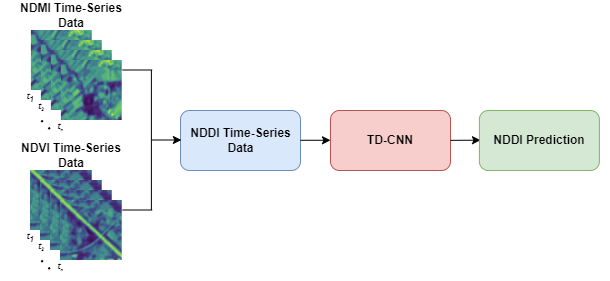}
    \end{subfigure}
    \begin{subfigure}[b]{.5\textwidth}
        \caption{Late NDDI Computation}
        \includegraphics[width = \textwidth]{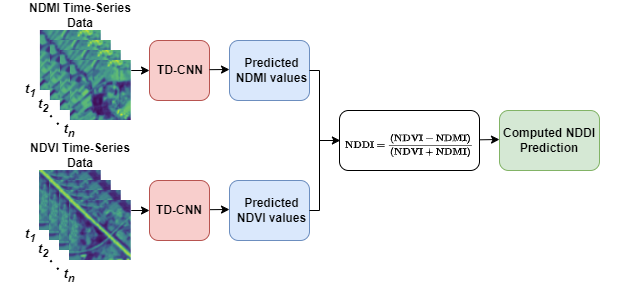}

    \end{subfigure}
    \caption{Early (a) and Late (b) NDDI Computation Workflow.}
    \label{workflowsearlylate}
    \vspace{-.4cm}
\end{figure}

\vspace{-1mm}
We explore two methods of computing NDDI values: early and late-stage computation. In \textbf{early-stage computation}, the NDVI and NDMI training and validation sets are combined using the formulas described in Eq. 1-3, to create a training and validation set of NDDI data, on which a single TD-CNN is trained. 
In \textbf{late-stage computation}, two separate TD-CNNs are trained on  NDMI and NDVI data respectively, and their predictions are used to compute a predicted NDDI frame, without training a model directly on NDDI data. 
Illustrations of early and late-stage computations are depicted in Figure \ref{workflowsearlylate}.

\begin{figure*}[!ht]
    \centering
    \begin{subfigure}[b]{.92\textwidth}
        \caption{NDMI Predictions (PR) and Ground Truths (GT)}
        \includegraphics[width = 
         \textwidth]{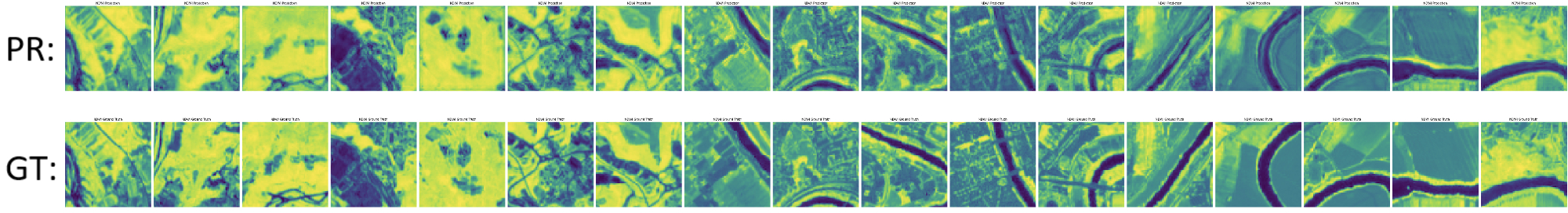}
    \end{subfigure}
    \begin{subfigure}[b]{.92\textwidth}
        \caption{ NDVI Predictions (PR) and Ground Truths (GT)}
        \includegraphics[width = 
         \textwidth]{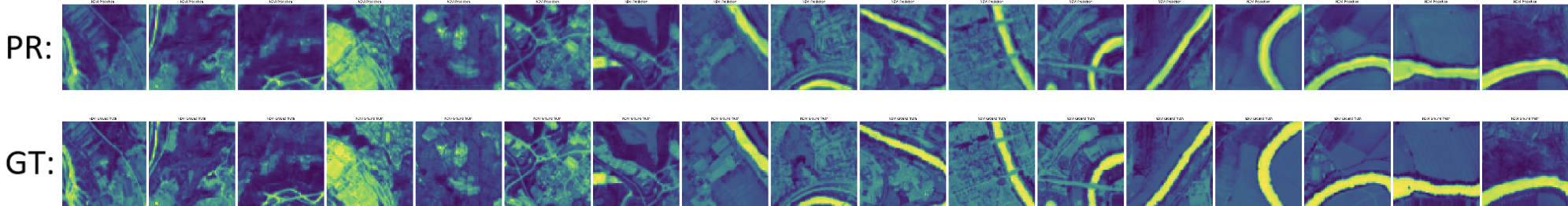}
    \end{subfigure}

    \begin{subfigure}[b]{.92\textwidth}
        \caption{ NDDI Late and Early Predictions (PR) and Ground Truths (GT)}
        \includegraphics[width = 
        \textwidth]{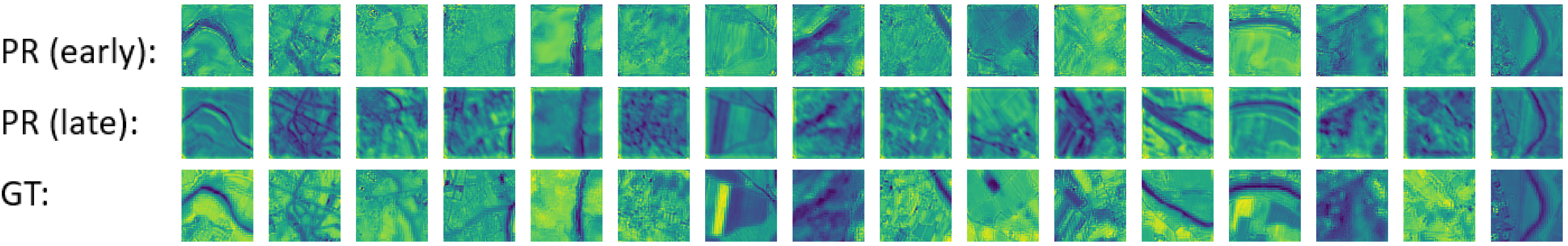}

    \end{subfigure}
    
    \caption{\centering NDMI, NDVI, and NDDI TD-CNN Prediction Results. For each (a), (b) and (c), the upper row(s) is a series of predictions for different locations, and the bottom row is a series of corresponding ground truths for each location.}
    \label{results}
    \vspace{-3mm}
\end{figure*}

\vspace{-0.35cm}
\section{RESULTS}
\vspace{-0.25cm}
The NDVI, NDMI, early-stage NDDI, and late-stage NDDI predictions are shown in Figure \ref{results}. Additionally, the Mean Squared Error (MSE), Structural Similarity Index (SSIM) and Peak Signal-to-Noise Ratio (PSNR) for the four prediction methods are summarized in Table 1. While the NDMI and NDVI indices prove to produce highly accurate next-frame prediction, NDDI experiences a noticeable loss in resolution. While this loss in resolution is somewhat typical with CNNs, it may indicate that NDDI is less suited towards long-term prediction than other indices. 

In particular, it seems that late NDDI prediction proves worse than early NDDI prediction. Considering that late NDDI prediction is computed using the outputs of two separate CNNs (NDVI and NDMI), it may be that this further detrimentally affects the resolution when compared to early NDDI prediction. 

In addition, SEN2DWATER, the dataset which was used to train models, uses satellite data taken from areas that are not drought-prone. It may be the case that NDDI is not a suitable index to describe drought or water conditions in non-drought-prone areas, and may perform better for predictive purposes if the model were trained on data taken from areas similar to those in \cite{Artikanur2022-ns}.

\begin{table}[!ht]
\renewcommand{\arraystretch}{1.1}
\begin{center}
\label{table-err-metrics}
\begin{tabular}{|l|c|c|c|}
\hline
\multicolumn{1}{|c|}{\textbf{Model}} & \multicolumn{1}{c|}{\textbf{MSE}} & \multicolumn{1}{c|}{\textbf{SSIM}} & \multicolumn{1}{c|}{\textbf{PSNR}} \\ \hline
\hspace{1.7mm}NDVI                                 & \hspace{1.7mm}0.00743\hspace{1.7mm}                           & 0.99885                            & 27.956                             \\
\hspace{1.7mm}NDMI                                 & 0.01065                           & 0.99967                            & 33.342                            \\
\hspace{1.7mm}NDDI (early)\hspace{1.7mm}                         & 0.00054                           & \hspace{1.7mm}0.99995\hspace{1.7mm}                            & \hspace{1.7mm}39.754\hspace{1.7mm}                             \\
\hspace{1.7mm}NDDI (late)                          & 0.0062                            & 0.99897                            & 28.440    \\                        
\hline
\end{tabular} 
\caption{Summary of Performance Metrics for Models. NDDI early computation outperforms late computation. }
\end{center}
\vspace{-9mm}
\end{table}

This work serves as a proof of concept for using the SEN2DWATER database for prediction via machine learning methods, and we propose several potential use cases for such models. The ability to potentially predict changes in moisture and water conditions is promising and directly useful for a wide variety of cases including agriculture and livestock production. Additionally, future work may look to implement such models to recover data when satellite captures of an area of land were taken during adverse weather conditions (clouds, fog) or when natural disasters (i.e. volcanic clouds) make it difficult to collect clear images. Additionally, future work may include multiple timestep predictions, should workarounds to loss of resolution be introduced.

\vspace{-0.35cm}
\section{CONCLUSIONS}
\vspace{-0.25cm}

In this paper, we present machine learning techniques as applied to the SEN2DWATER dataset for the purposes of drought prediction. 
We train TD-CNNs, on three indices,  NDMI, NDVI, and NDDI, and find that NDMI and NDVI maintain high accuracy for next-frame prediction. While NDDI produces reasonable predictions during early computation, this index displays significant loss of resolution in  late computation. For NDDI to be a practical index to use in learning-based prediction methods, future research could work to improve NDDI prediction by minimizing loss of resolution. Alternatively, other  indices could be explored that encode similar information as the NDDI but behave better when learned on by existing time-distributed models, such as the Standardized Precipitation Evapotranspiration Index (SPEI).

\vspace{-0.35cm}
\section{Acknowledgments}
\vspace{-0.22cm}

\noindent 
We would like to acknowledge MIT's MISTI research exchange program for making this scientific collaboration possible. The SEN2DWATER dataset was created by students at the University of Sannio, and the works presented here are contributions from Veronica Muriga and Benjamin Rich from MIT, and Francesco Mauro, Alessandro Sebastianelli, and Silvia Liberata Ullo from University of Sannio.

\vspace{-0.45cm}
\bibliographystyle{IEEEbib}
\bibliography{refs}

\end{document}